\documentclass[amssymb,amsmath,aps,showpacs,floatfix,nofootinbib,showpacs,12pt]{revtex4}
\usepackage{graphicx,amssymb,color,soul,latexsym}

\begin{document}

\title{Perspective of Galactic dark matter subhalo detection on Fermi
from the EGRET observation\footnote{Supported by National Natural
Science Foundation of China (10575111, 10773011) and Chinese Academy
of Sciences (KJCX3-SYW-N2).} }

\author{
YUAN Qiang$^{1}$\footnote{yuanq@mail.ihep.ac.cn}, BI Xiao-Jun$^{1,2}$
and ZHANG Juan$^{1}$}

\affiliation{$^{1}$ Key Laboratory of
Particle Astrophysics, Institute of High Energy Physics, Chinese
Academy of Sciences, Beijing 100049, P. R. China \\
$^{2}$ Center for High Energy Physics,
Peking University, Beijing 100871, P.R. China
}

\begin{abstract}
The perspective of the detectability of Galactic dark matter
subhaloes on the Fermi satellite is investigated in this work. Under
the assumptions that dark matter annihilation accounts for the ``GeV
excess'' of the Galactic diffuse $\gamma$-rays discovered by EGRET
and the $\gamma$-ray flux is dominated by the contribution from
subhaloes of dark matter, we calculate the expected number of dark
matter subhaloes that Fermi may detect. We show that Fermi may
detect a few tens to several hundred subhaloes in 1-year all sky
survey. Since EGRET observation is taken as a normalization, this
prediction is independent of the particle physics property of dark
matter. The uncertainties of the prediction are discussed in detail.
We find that the major uncertainty comes from the mass function of
subhaloes, i.e., whether the subhaloes are ``point like'' (high-mass
rich) or ``diffuse like'' (low-mass rich). Other uncertainties like
the background estimation and the observational errors will
contribute a factor of $2\sim 3$.
\end{abstract}


\maketitle

\section{Introduction}

The dark matter (DM) problem is one of the most important issues in
modern physics and cosmology. Unfortunately, more than seventy years
after the discovery of DM, the nature of the DM particle is still
unclear. To identify the DM particles, it is necessary to ``see''
them in particle physics experiments beyond the gravitational
measurements. There are usually three types of experiments suggested
to capture the DM particles. The first one is to produce DM particle
pairs by high energy particle collision at colliders such as the
forthcoming Large Hadron Collider (LHC,
\cite{Drees:2000he,Feng:2005gj,Baer:2008uu,Kane:2008gb}). The second
method is the so-called {\it direct search} of DM particle by
looking for the signals when DM scattering off the detector nuclei
\cite{Bednyakov:1994te,Bernabei:1998td,Munoz:2003gx,Trotta:2006ew},
which is thought to be the most direct way to show the existence and
understand the property of DM particle. Finally the {\it indirect
search} of DM annihilation products in cosmic rays (CRs) including
$\gamma$-rays \cite{Bouquet:1989sr,
Bergstrom:2001jj,Elsaesser:2004ap,deBoer:2006tv}, anti-particles
\cite{Baltz:2001ir,Donato:2003xg,Donato:2008yx} and neutrinos
\cite{Albuquerque:2000rk,Barger:2001ur,Halzen:2005ar,
Barger:2007hj}, is also an important complementary method for DM
searches.

Among the annihilation products, $\gamma$-ray is the most attractive
one for detection. Comparing with neutrinos, $\gamma$-rays are
easier to be recorded; while comparing with charged anti-particles,
$\gamma$-rays will not be deflected by the magnetic field and can
trace back to the sites where the annihilation takes place. In this
work we focus on the $\gamma$-rays from DM annihilation.

It is known that DM annihilation products are proportional to the
density square of DM distributions. Therefore the highly
concentrated regions are good sites for DM searches. Theoretically
there are many such sites proposed to search for DM, such as the
Galactic center (e.g.,
\cite{Gondolo:1999ef,Merritt:2002vj,Cesarini:2003nr,
Dodelson:2007gd}), DM substructures \cite{Bi:2005im,Bi:2007pr,
Diemand:2006ik,Yuan:2006ju,Pieri:2007ir}, dwarf galaxies
\cite{Hooper:2003sh,Bi:2006av,Colafrancesco:2006he,Strigari:2006rd},
and the mini-spikes
\cite{Bertone:2005xz,Fornasa:2007ap,Brun:2007tn}. It has been
recognized that DM subhaloes have several advantages for DM searches
when comparing with other sites \cite{Bi:2007pr,Wai:2007zza}.
Firstly, the subhaloes distribute isotropically in the Galactic
halo, and can easily be located in a low background environment away
from the Galactic plane. Secondly, the subhalo may well decouple
from baryon matter and give clean DM signals. Thirdly, large number
of subhaloes makes it possible to do statistical study. Therefore to
draw attention to the signals from DM subhaloes should be a very
important aspect for DM {\it indirect searches}.

From the observational point of view, the Energetic Gamma Ray
Experiment Telescope (EGRET) on board the satellite Compton Gamma
Ray Observatory (CGRO) surveyed the high energy $\gamma$-ray sky in
$30$ MeV$-30$ GeV. EGRET discovered $271$ $\gamma$-ray sources
\cite{Hartman:1999fc} and measured the all-sky diffuse $\gamma$-ray
emission \cite{Hunger:1997we,Cillis:2005bd}. Among the $271$ sources
of the third EGRET catalog, only $101$ ones are identified as known
sources. There are $48$ additional sources that are suggested to be
associated with possible counterparts. A substantial fraction with
over $120$ sources is still unidentified. The DM subhaloes are
thought to be possibly composed of one population of such
unidentified sources. However, the analysis of the luminosity and
spatial distributions show that no more than $\sim20$ sources can be
attributed to DM clumps \cite{Flix:2004ks,SiegalGaskins:2007dx}.
While for the Galactic diffuse $\gamma$-ray emission, the spectrum
shows an ``excess'' at energy $\sim$ GeV compared with that
predicted by the conventional CR model \cite{Hunger:1997we}. An
appealing interpretation of the ``GeV excess'' is DM annihilation
\cite{deBoer:2005tm,deBoer:2004es, deBoer:2007sq}. By fitting the
all sky diffuse $\gamma$-ray distributions, de Boer et al. found
that the all-sky spectra can be well reproduced by $50-70$ GeV
supersymmetric DM annihilation, with an isothermal halo distribution
and two additional ring-like structures \cite{deBoer:2005tm}.
However, a ``boost factor'' of $\sim 100$ is needed to match the
absolute fluxes. The ``boost factor'' is proposed to be able to come
from the enhancement of annihilation rate by DM subhaloes. Bi et al.
have constructed a realistic model to explain both the diffuse
$\gamma$-ray spectra and the CR observations, taking into account
the DM subhaloes from numerical simulations
\cite{Bi:2006vk,Bi:2007ab}. It is found that a strongly clumpy DM
profile is needed to give such a large ``boost factor''. In order to
further test this scenario we have to turn to the experiments with
higher sensitivity, such as the newly launched Fermi
satellite\footnote{http://glast.gsfc.nasa.gov/}, which will follow
the footsteps of CGRO-EGRET to explore the high energy $\gamma$-ray
sky. The Fermi-LAT (Large Area Telescope) is a high quality
instrument with larger field-of-view ($\sim2.5$ sr), wider energy
bands ($10$ MeV$-300$ GeV) and higher sensitivity compared with
EGRET. It will be able to improve the sensitivity of {\it indirect
search} of DM annihilation greatly.

Considering the advantages of detecting DM annihilation from the
Galactic subhaloes, it is now one of the most prominent scientific
goals to search for DM subhaloes by Fermi. The perspective of DM
detection from subhaloes on the Fermi satellite has been discussed
extensively in literature (e.g., \cite{Peirani:2004wy,Wai:2007px,
Kuhlen:2008aw,Baltz:2008wd}). Assuming a specific DM model, Baltz et
al. claimed that for 5-year exposure Fermi might detect $\sim 10$ DM
satellites (subhaloes) with significance $>5\sigma$
\cite{Baltz:2008wd}. Kuhlen et al. found that based on the numerical
simulation of DM distributions the number of subhaloes Fermi might
detect was from a few to several dozens depending on the subhalo
boost factor and DM annihilation parameters \cite{Kuhlen:2008aw}.
Bertone et al. showed that DM annihilation around intermediate mass
black holes (mini-spikes) might be a population of bright
$\gamma$-ray sources of Fermi \cite{Bertone:2005xz}. However, all of
these kinds of studies depend sensitively on the particle physics
model of DM particles and suffer large uncertainties.

In the present work we try to investigate the performance of DM
subhalo detection on Fermi, under the assumptions that the EGRET
``GeV excess'' comes from DM annihilation and the DM induced
$\gamma$-ray emission is dominated by that from the subhaloes. Since
the EGRET observation of diffuse $\gamma$-rays is adopted as a
normalization, this prediction is expected to get rid of the
uncertainties from the annihilation cross section and $\gamma$-ray
yield spectrum. The degeneracy between the normalization factor of
DM subhaloes and the annihilation cross section is discussed.
Furthermore we discuss the uncertainties of this prediction.

The outline of the paper is as follows. We first introduce the model
of DM annihilation scenario to explain the EGRET ``GeV excess'' in
Sec.~2. In Sec.~3 we give the prediction of detectable number of
subhaloes on Fermi based on this model. The uncertainties of the
prediction are carefully discussed in Sec.~4. Finally we give
conclusion and discussion in Sec.~5.

\section{The model to account for the EGRET observations}

\begin{figure}[!htb]
\begin{center}
\includegraphics[width=0.8\columnwidth]{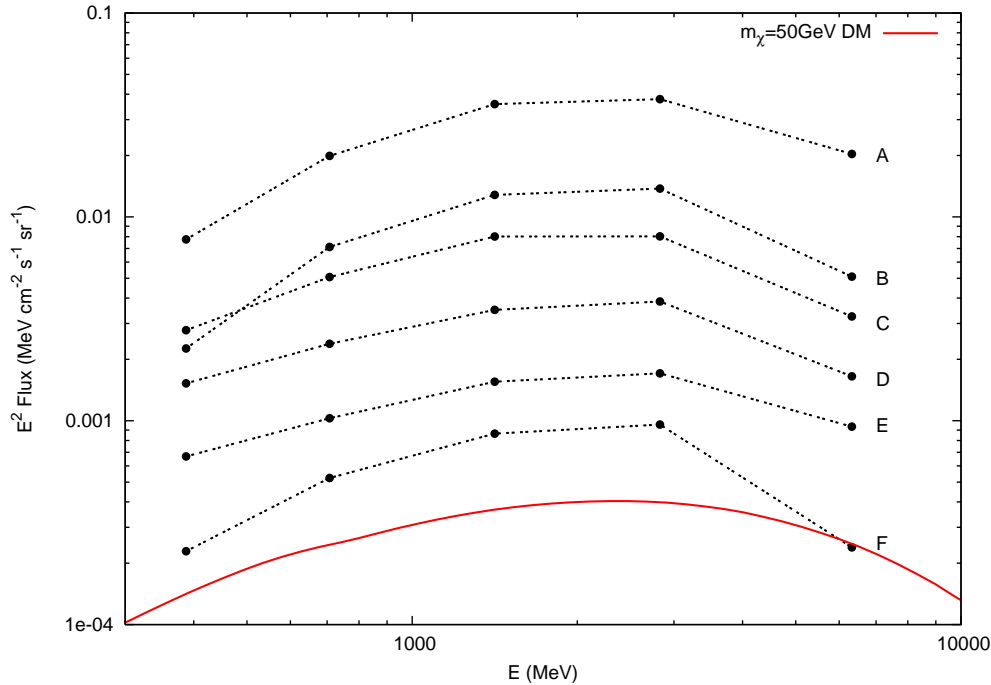}
\caption{The differences between the observational EGRET fluxes
and the CR induced $\gamma$-rays predicted from GALPROP. Similar
shapes of different sky regions indicate a common origin of the
excess. The solid line shows the DM annihilation spectrum (with
arbitrary normalization) for neutralino with mass $50$ GeV.}
\label{fig1}
\end{center}
\end{figure}

One way to explain the EGRET ``GeV excess'' is the DM annihilation
scenario \cite{deBoer:2005tm,deBoer:2004es,deBoer:2007sq}. This
scenario has been extended to be a more realistic one to explain
both the diffuse $\gamma$-ray and CR observations
\cite{Bi:2006vk,Bi:2007ab}. For the convenience of discussion in
this work, we base on the model of Ref.\cite{Bi:2007ab}, however,
the conclusion will not be limited in this detailed model. In the
model of Ref.\cite{Bi:2007ab}, the background $\gamma$-rays are
calculated using the package GALPROP
\cite{Strong:1998pw,Strong:1998fr}. The parameters of GALPROP are
adjusted to reproduce the EGRET data and the CR observations,
especially the antiproton data, after including the contribution
from DM annihilation. In Fig.\ref{fig1} we show the differences
between the observational spectra by EGRET and the predicted
background (including the extra-galactic component) at different sky
regions. The six sky regions are defined following Ref.
\cite{Strong:2004de}: region A corresponds to the inner Galaxy with
$0^{\circ}<l<30^{\circ},\ 330^{\circ}<l<360^{\circ}$ and
$|b|<5^{\circ}$; region B is the Galactic plane excluding region A
with $30^{\circ}<l< 330^{\circ}$ and $|b|<5^{\circ}$; C is the outer
Galaxy with $90^{\circ} <l<270^{\circ}$ and $|b|<10^{\circ}$;
regions D, E, F cover the medium and high latitudes at all
longitudes with $10^{\circ}<|b|<20^{\circ}$,
$20^{\circ}<|b|<60^{\circ}$ and $60^{\circ}<|b|<90^{\circ}$
respectively. From the figure we notice that the differences of the
spectra at different regions are very similar, which implies that
there may be some common origins of the lacked $\gamma$-rays, for
example the DM annihilation
\cite{deBoer:2005tm,deBoer:2004es,deBoer:2007sq}.

The diffuse $\gamma$-ray fluxes from DM annihilation are determined
by two factors: the ``particle factor'' which describes the particle
physics property of DM particles and the ``astrophysics factor''
which describes the spatial distribution of DM. In Ref.
\cite{Bi:2007ab} the ``particle factor'' is calculated under the
framework of the minimal supersymmetric extension of the standard
model (MSSM, \cite{Jungman:1995df}). The lightest particle
neutralino in MSSM is adopted as the DM particle \footnote{However,
it will be shown in the following that this assumption can be
relaxed in the present work.}. The annihilation cross section and
$\gamma$-ray production spectrum are calculated using DarkSUSY
\footnote{http://www.physto.se/~edsjo/darksusy/}, a package for MSSM
DM calculation \cite{Gondolo:2004sc}. A random scan in the MSSM
parameter space is performed and the adopted models are chosen to
satisfy the relic density constraint and give large $\gamma$-ray
fluxes. The annihilation spectrum for the neutralino with mass
$m_{\chi}\approx 50$ GeV, as shown in Fig.\ref{fig1}, is found to be
in good agreement with the observations.

The ``astrophysics factor'' is mainly determined according to the
numerical simulation of DM distribution. High resolution simulations
show that a fraction of $10\%\sim 20\%$ of the total mass survives
in the self-bound substructures
\cite{tormen98,Moore:1999nt,Ghigna:1999sn,Zentner:2003yd,
DeLucia:2003xe,Diemand:2005vz}. The substructures of DM can
effectively enhance the annihilation signals
\cite{Bi:2005im,Yuan:2006ju}. The number density distribution of
subhaloes from simulation can be fitted by an isothermal spatial
distribution and a power-law mass function
\begin{equation}
\frac{{\rm d}N}{{\rm d}m_{\rm sub}\cdot4\pi r^2{\rm d}r}=
N_0\left(\frac{m_{sub}}{M_{\rm vir}}\right)^{-\alpha}
\frac{1}{1+\left(r/r_{\rm H}\right)^2}, \label{number}
\end{equation}
where $M_{\rm vir}\approx 10^{12}$ M$_{\odot}$ is the virial mass of
the Galaxy, $r_{\rm H}\approx 0.14r_{\rm vir}=29$kpc is the core
radius of the distribution of subhaloes \cite{Diemand:2004kx}, and
$N_0$ is the normalization factor determined by the total mass of
subhaloes. The slope $\alpha$ varies from $1.7$ to $2.1$ in various
works
\cite{Moore:1999nt,Ghigna:1999sn,Helmi:2002ss,DeLucia:2003xe,Gao:2004au,
Shaw:2005dy}. The intermediate value $\alpha\approx 1.9$ is favored
by the recent highest resolution simulation Via Lactea
\cite{Madau08}. In Ref.\cite{Bi:2007ab} $\alpha=1.9$ is adopted. The
masses of subhaloes range from the minimum $\sim 10^{-6}$
M$_{\odot}$ which is close to the free-streaming mass
\cite{Diemand:2005vz,Hofmann:2001bi,Chen:2001jz} to the maximum
about $0.01M_{\rm vir}$. The DM density distribution inside the
subhalo follows a $\gamma$ profile $\rho(r)=\frac{\rho_{\rm
s}}{(r/r_{\rm s})^{\gamma} (1+r/r_{\rm s})^{3-\gamma}}$ with
$\gamma=1.7$, which is steeper in the center than the NFW type
($\propto r^{-1}$, \cite{Navarro:1996gj}) and Moore type ($\propto
r^{-1.5}$, \cite{Moore:1999gc}). A steeper inner slope for small
mass halo is suggested by numerical simulations
\cite{Reed:2003hp,Diemand:2006ey}. The profile parameters $r_{\rm
s}$ and $\rho_{\rm s}$ are determined according to the halo mass
$M_{\rm vir}$ and the concentration parameter $c_{\rm vir}$
\cite{Bullock:1999he,Yin:2008mv}. The $c_{\rm vir}-M_{\rm vir}$
relation is adopted as the model of Bullock et al.
\cite{Bullock:1999he}. Finally the central density of a subhalo is
truncated at a certain radius to avoid the divergence. A
characteristic radius $r_{\rm cut}$ is introduced within which the
DM density is kept a constant $\rho_{\rm max}$ due to the balance
between the annihilation rate and the in-falling rate of DM
\cite{Berezinsky:1992mx}. Typically we have $\rho_{\rm max}=
10^{18}\sim10^{19}$ M$_{\odot}$ kpc$^{-3}$ \cite{Lavalle:1900wn}.

This configuration of DM distribution is then used to calculate the
diffuse $\gamma$-ray emission. Taking $\rho_{\rm max}= 10^{19}$
M$_{\odot}$ kpc$^{-3}$ and subhalo mass fraction $20\%$
\footnote{Note that the parameter adoption is a bit different from
Ref.\cite{Bi:2007ab}, where $r_{\rm cut}$ instead of $\rho_{\rm
max}$ is used.} we find that the results can well reproduce the high
latitude observations by EGRET (i.e., F region). It can be noted
from Fig.2 of Ref.\cite{Bi:2007ab} that for the direction far away
from the Galactic center the contribution to $\gamma$-rays from
subhaloes dominates that from the smooth halo. For example at the
Galactic pole direction $b=\pm 90^{\circ}$, the ratio of the
astrophysics factors between the subhaloes and smooth halo is
$f=\Phi_{\rm sub}/\Phi_{\rm sm} \approx10^2$, which is consistent
with the requirement of a boost factor $\sim 100$ of
Ref.\cite{deBoer:2005tm}. For other sky regions especially for the
Galactic plane regions, the enhancement of subhaloes is not enough,
and two additional DM rings are needed \cite{deBoer:2005tm}. The
overall results including the DM rings are well consistent with the
observations.

\section{Detection performance of subhaloes on Fermi}

\begin{figure}[!htb]
\begin{center}
\includegraphics[width=0.8\columnwidth]{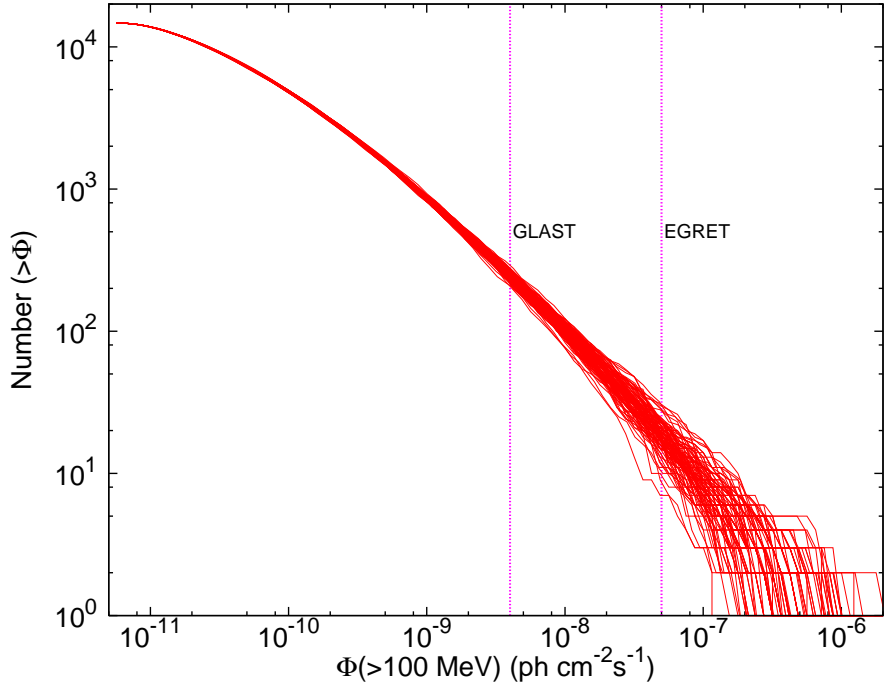}
\caption{The accumulative number of subhaloes as a function of
integral flux for energy threshold $E_{\rm th}=100$ MeV. The two
vertical lines show the sensitivities of EGRET and Fermi.}
\label{fig2}
\end{center}
\end{figure}

In this section we calculate the $\gamma$-ray fluxes from massive DM
subhaloes adopting the same DM model presented in Sec. 2 and discuss
the detectability of these $\gamma$-ray sources on Fermi. The
Monte$-$Carlo (MC) realization method is adopted to generate the DM
subhaloes with mass $\gtrsim 10^6$ M$_{\odot}$ in the Milky Way,
following the distribution function Eq.(\ref{number}). There are
about $1.5\times 10^4$ subhaloes found with mass heavier than $10^6$
M$_{\odot}$. Totally $100$ Milky Way like galaxies are generated.
For each realization, we calculate the annihilation flux of each
subhalo with energy threshold $E_{\rm th}=100$ MeV and count the
accumulative number as a function of the threshold flux $\Phi$. The
result is shown in Fig.\ref{fig2}.

\begin{figure}[!htb]
\begin{center}
\includegraphics[width=0.45\columnwidth]{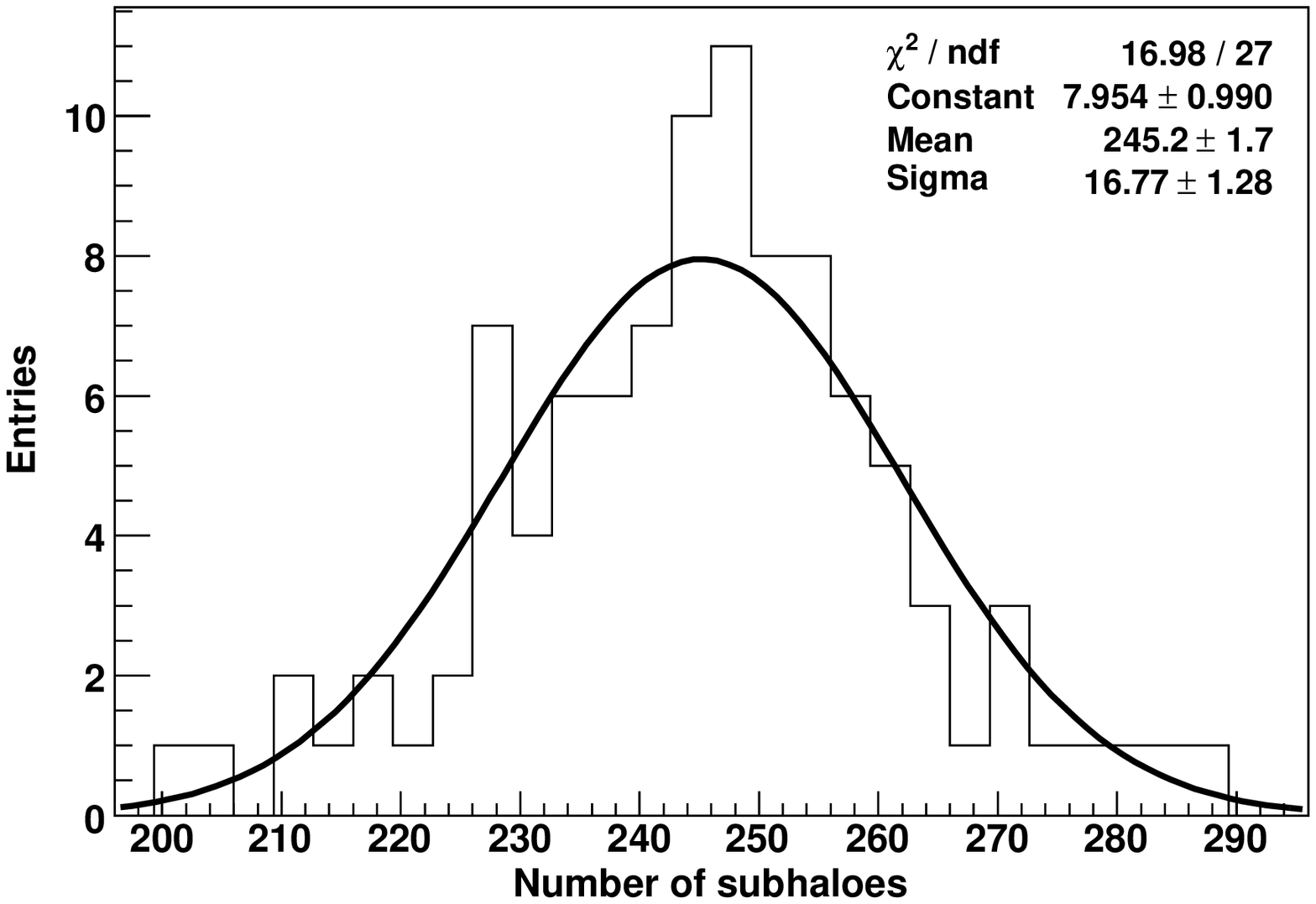}
\includegraphics[width=0.45\columnwidth]{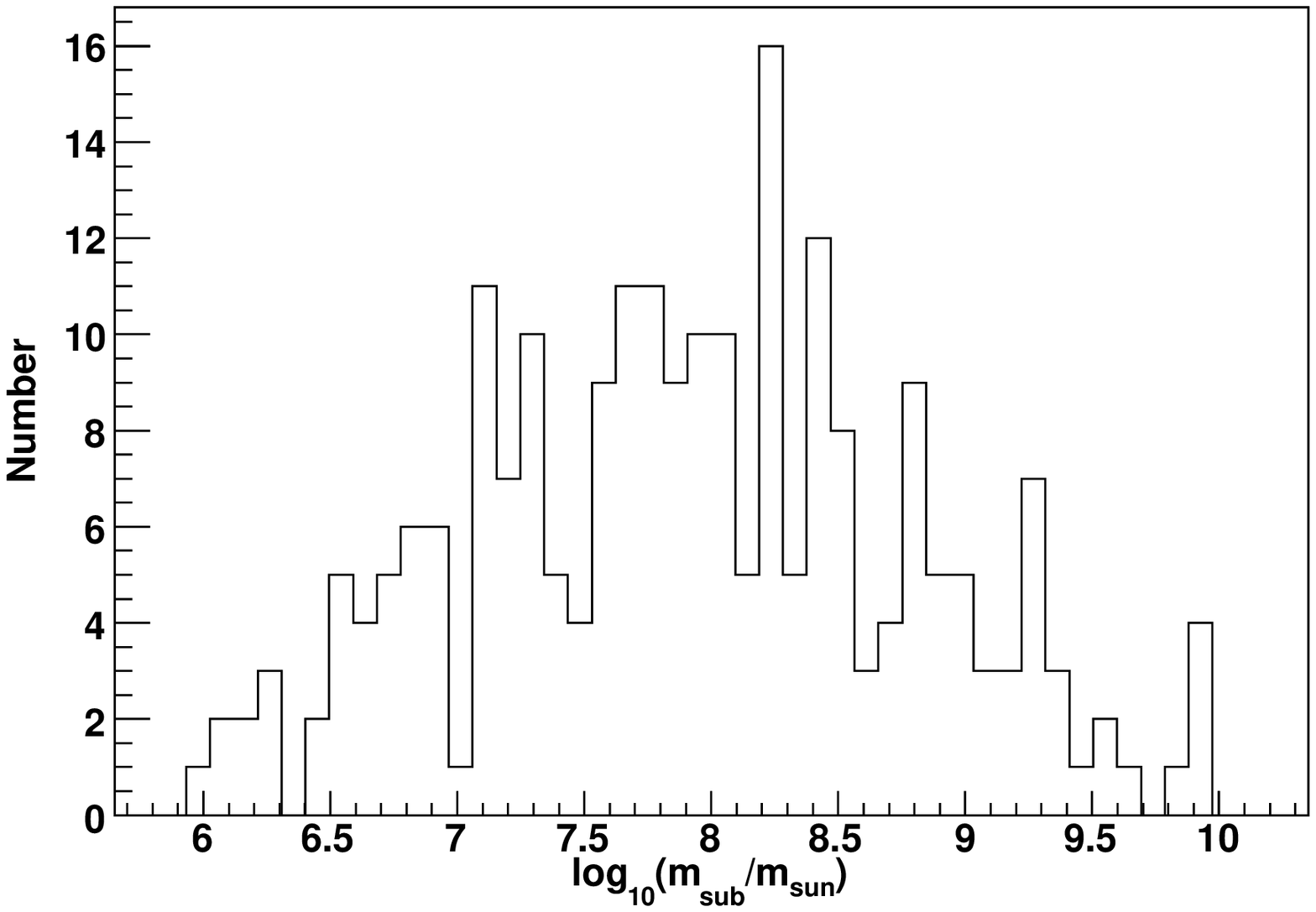}
\includegraphics[width=0.45\columnwidth]{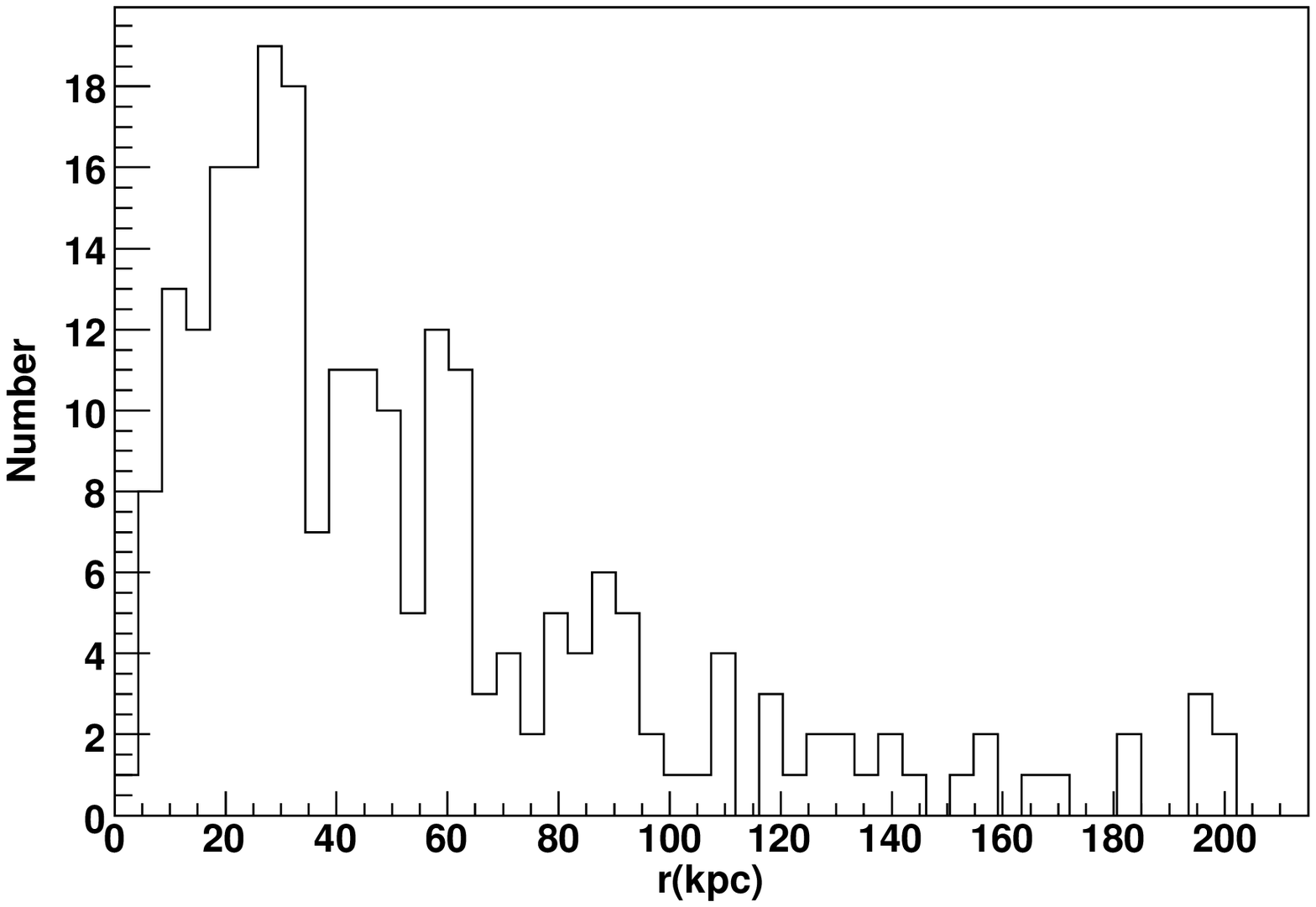}
\includegraphics[width=0.45\columnwidth]{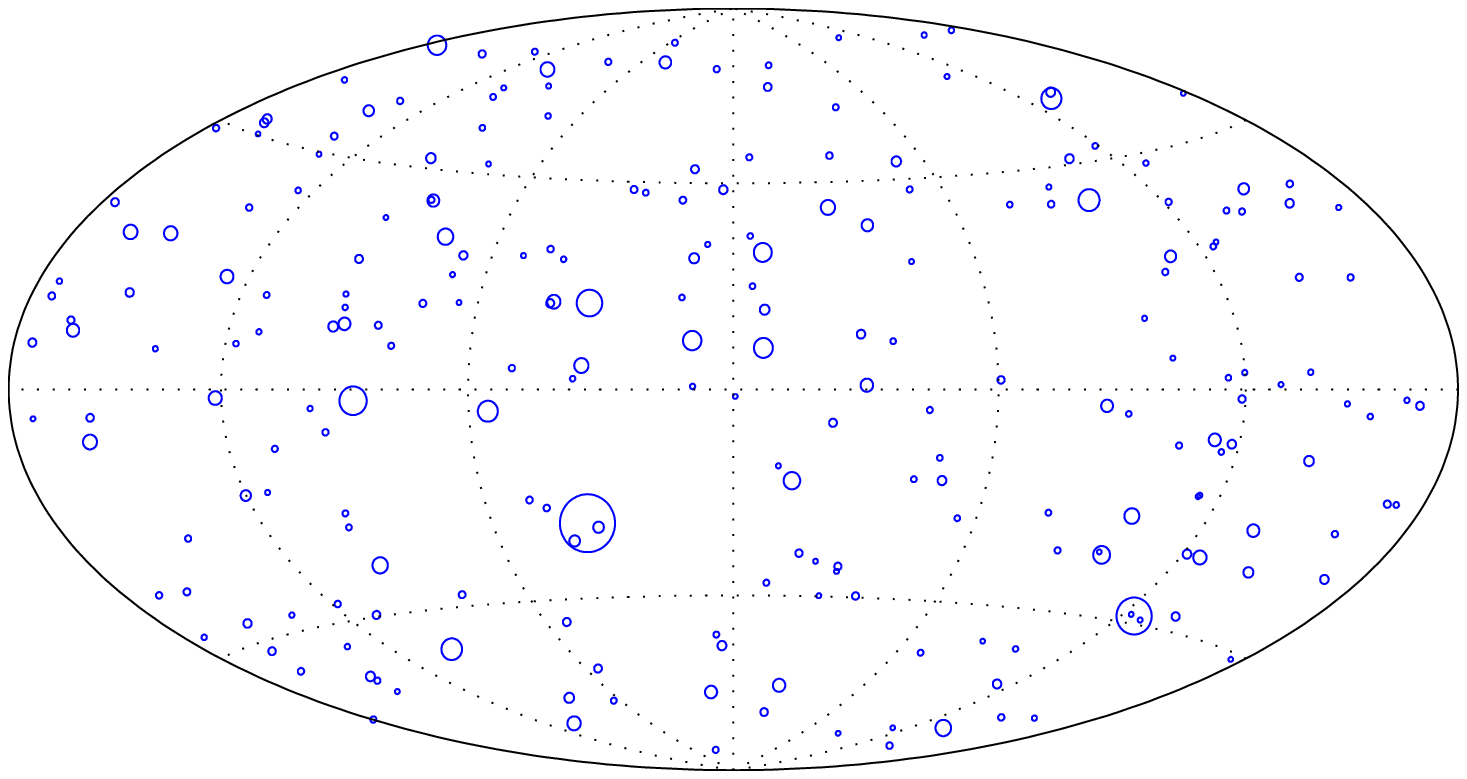}
\caption{Distributions of the detectable DM subhaloes on Fermi:
detectable number distribution due to various realizations ({\it
top-left}); mass ({\it top-right}), distance ({\it bottom-left})
distributions and skymap ({\it bottom-right}) of the detectable DM
subhaloes in one realization. The area of the circle in the skymap
is proportional to the flux.} \label{fig3}
\end{center}
\end{figure}

The sensitivities of EGRET and Fermi for $1$-year all-sky survey at
$5\sigma$ are $5\times10^{-8}$ ph cm$^{-2}$ s$^{-1}$ ($>100$ MeV)
\cite{Morselli:2002cc} and $4\times 10^{-9}$ ph cm$^{-2}$ s$^{-1}$
($>100$ MeV) respectively\footnote{
http://www-Fermi.slac.stanford.edu/software/IS/glast\_lat\_performance.htm},
which are shown by the vertical lines in Fig.\ref{fig2}. From the
figure we find that the number of subhaloes that EGRET can detect is
$18.7\pm 4.4$, which is consistent with the results of
Refs.\cite{Flix:2004ks,SiegalGaskins:2007dx}. For Fermi, the
detectable number is $245.2\pm 16.8$, which is an order of magnitude
more than EGRET. The scattering comes from different realizations.
The probability distribution of the detectable numbers on Fermi is
well fitted with a Gaussian distribution, as shown in the {\it
top-left} panel of Fig.\ref{fig3}. We also plot the mass
($\log_{10}[m_{\rm sub}/M_{\odot}]$) distribution, the distance
distribution and the directional skymap in Galactic coordinate of
the detectable DM subhaloes on Fermi for one of the realizations in
Fig.\ref{fig3}. It is shown that subhaloes with masses $\sim 10^8$
M$_{\odot}$ and distances within $\sim 50$ kpc are more likely to be
detected by Fermi. The direction distribution is isotropic, which
will be significantly different from other astrophysical populations
of sources.

\begin{figure}[!htb]
\begin{center}
\includegraphics[width=0.8\columnwidth]{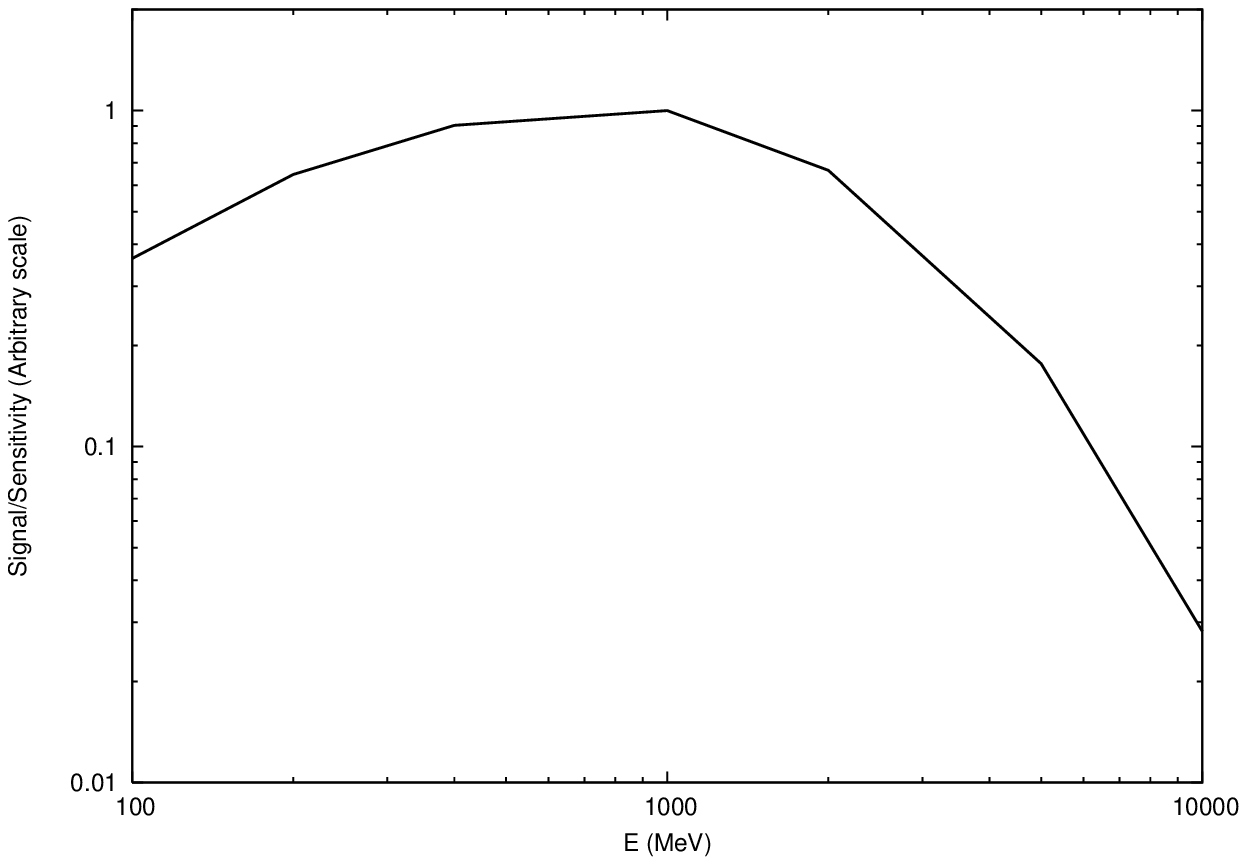}
\caption{Integral signal from DM annihilation to sensitivity of
Fermi ratio as a function of threshold energy. }\label{fig4}
\end{center}
\end{figure}

In Fig.\ref{fig2} the threshold energy of the detector is adopted as
$100$ MeV. However, since the energy spectrum of DM annihilation is
known, as shown in Fig.\ref{fig1}, the detector performance can be
optimized by taking a proper energy cut. In Fig.\ref{fig4} we show
the ratio between the integral flux of DM annihilation spectrum and
the integral sensitivity of Fermi. It will be more efficient for the
DM detection for larger ratio. It can be seen that for the energy
threshold $E_{\rm th} \approx 1$ GeV, the detectability is most
optimized. We show that for $E_{\rm th}= 1$ GeV the number of
subhaloes that can be detected by Fermi is $620.4\pm 23.1$. In the
following of this paper we will adopt the optimized threshold energy
$E_{\rm th}= 1$ GeV for discussion.

\section{Uncertainties}

In the previous section, we show our prediction of the detectable
number of DM subhaloes on Fermi based on the EGRET observations of
the Galactic diffuse $\gamma$-rays. It is about $250$ (or $620$ for
optimized energy cut) according to the model of
Ref.\cite{Bi:2007ab}. In this section we discuss the possible
uncertainties of this prediction.

\subsection{Particle factor}

There are two quantities in the particle factor to affect the
annihilation flux: the velocity weighted averaged cross section
$\langle\sigma v\rangle$ and the $\gamma$-ray production spectrum
per annihilation ${\rm d}N/{\rm d}E$. In
Refs.\cite{deBoer:2005tm,Bi:2007ab}, the neutralino is adopted as
the DM particle. $\langle\sigma v\rangle$ and ${\rm d}N/{\rm d}E$
are calculated under the MSSM model. However, from Fig.\ref{fig1} it
can be seen that the annihilation spectrum is fixed by the ``gaps''
between the observations and the background, and is independent with
the DM particle models. We find that if we adopt a model independent
spectrum by averaging the spectra of the $6$ sky regions (normalized
by the maximum value and extrapolate logarithmically to higher and
lower energies) instead of the spectrum from neutralino
annihilation, the results are almost the same.

The annihilation cross section $\langle\sigma v\rangle$ will affect
the absolute flux of $\gamma$-rays. Since the absolute flux is
determined by the EGRET observation, there is degeneracy between the
cross section and the astrophysics factor. That is to say, a smaller
cross section $\langle\sigma v\rangle$ needs to be compensated by a
larger astrophysics factor. Therefore, if the total number of DM
subhaloes maintains the same, varying $\langle\sigma v\rangle$ is
not expected to change the detectability of DM subhaloes
significantly. We will show in the next section that even the
normalization of the total number changes, the conclusion still
holds. However, it should be pointed out that, if $\langle\sigma
v\rangle$ is so large that the smooth contribution dominates the
$\gamma$-ray emission from DM annihilation (i.e., $f=\Phi_{\rm
sub}/\Phi_{\rm sm}\lesssim 1$), it will be more difficult to detect
DM subhaloes.

\subsection{Astrophysics factor}

In the astrophysics factor, the normalization of number of
subhaloes, the mass function of subhalo number distribution, the
inner profile of DM subhalo and the concentration model may all lead
to uncertainties of the predicted detectable number of subhaloes
given in Sec. III. We now investigate these issues one by one in
detail.

The normalization of the total number of subhaloes, i.e., $N_0$ of
Eq.(\ref{number}), is determined by the mass fraction of clumps in
the Galactic halo. It shows relatively large uncertainty in
different simulations. However, we find that even though the factor
$N_0$ is left free in the range of one or two orders of magnitude,
the result of this work is still unchanged. The reason is that in
order to keep the $\gamma$-ray emission of region F unchanged,
varying $N_0$ by some factor needs to be compensated by the same
factor of the annihilation flux of each subhalo (e.g., through the
rescale of $\langle\sigma v\rangle$). From Fig.\ref{fig2} we can see
that the cumulative number $N(>\Phi)\propto\Phi^{-1}$, which means
that the change of $N_0$ is equivalent to shifting the curves in
Fig.\ref{fig2} downward (or upward) while the change of
$\langle\sigma v\rangle$ shifts the curves rightward (or leftward)
by the same factor. Therefore the number of subhaloes with flux
higher than the sensitivity of Fermi will keep the same.

The index of the mass function $\alpha$ in Eq. (\ref{number}) will
affect the mass distribution of subhaloes, and accordingly affect
the $N(>\Phi)\sim\Phi$ relation. If $\alpha$ is smaller, the
fraction of contribution to the diffuse $\gamma$-rays from high mass
subhaloes becomes more important (see Fig.4 of
Ref.\cite{Lavalle:1900wn}), so we can expect that the subhaloes with
higher fluxes are richer. The expected numbers of subhaloes on Fermi
for energy threshold $1$ GeV, for different $\alpha$ are listed in
Table \ref{table1}. For each case, we scale $\langle\sigma v\rangle$
to recover the EGRET observation at region F. The inner DM profile
$\gamma=1.7$, the central maximum density $\rho_{\rm max}=10^{19}$
M$_{\odot}$ kpc$^{-3}$ and subhalo mass fraction $20\%$ are kept
unchanged \footnote{The same treatment is also employed in the
following discussion, i.e., we rescale $\langle\sigma v\rangle$ to
cancel the effect by the relevant changes and keep other settings
unchanged as the benchmark model presented in Sec. 2.}. It is shown
that for different values of $\alpha$ the results differ from each
other significantly. The parameter $\alpha$ determines the DM to be
high mass rich (``point like'') or low mass rich (``diffuse like''),
which is the key factor for the detection of DM subhaloes. It should
be noted that for the cases $\alpha=1.7$ and $1.8$, the most
luminous DM subhalo is even brighter than the brightest unidentified
EGRET source \cite{Hartman:1999fc}. In order not to break the
constraint from the EGRET source catalog we find $\alpha\gtrsim
1.9$. The index of the mass function of subhaloes $\alpha$ is
actually the most important source of the uncertainties when
predicting the detectability of DM annihilation from subhaloes.

\begin{table}[htb]
\begin{center}
\caption{\label{table1}Detectable number of DM subhaloes on Fermi
for energy threshold $1$ GeV.}
\begin{tabular*}{120mm}{@{\extracolsep{\fill}}ccccccc}
\hline \hline
$\alpha$ & $1.7$ & $1.8$ & $1.9$ & $2.0$ & $2.1$ \\
\hline
 Number & $4068\pm32.8$ & $2147\pm40$ & $620.4\pm23.1$ & $110.9\pm10.5$ &
$13.6\pm3.6$ \\
\hline\hline
\end{tabular*}
\end{center}
\end{table}

The inner slope of DM subhalo can also result in slightly different
mass dependence of the annihilation flux, which may also change the
detection performance of subhaloes on Fermi. This effect is not very
important. We show that for Moore profile, $N_{\rm det}=410.8\pm
20.0$, and for NFW profile, $N_{\rm det}=441.4\pm 20.3$. From Fig.3
of Ref.\cite{Yuan:2006ju} we can see that for $\gamma=1.7$ profile
the fraction of DM annihilation luminosity from high mass subhaloes
is more important than that of Moore or NFW profiles. Therefore the
detectable number $620.4\pm 23.1$ for $\gamma=1.7$ profile (i.e.,
the benchmark model in this work), is some larger than that of NFW
or Moore.

Finally, the concentration model adopted in Ref.\cite{Bi:2007ab} is
from Ref.\cite{Bullock:1999he}. At present there is no good
determination about the subhalo concentration, especially for the
low mass haloes which are beyond the resolution of simulations. As a
comparison, using another concentration model of Eke et al.
\cite{Eke:2000av} we find $N_{\rm det}=794.6\pm 26.3$. The model of
Eke et al. \cite{Eke:2000av} gives larger concentration (accordingly
larger annihilation flux) of heavier halo, resulting in more
detectable number of subhaloes.

\subsection{Background estimation}

The background $\gamma$-ray emission from CRs is calculated based on
the conventional CR propagation model. To discuss the uncertainty of
the background calculation is very complicated due to the
uncertainties of the inputs, including gas distribution in the
Galaxy, interstellar radiation field, nuclear interaction cross
section, CR measurements, CR source distribution, solar modulation
and so on \cite{Strong:2004de}. However, many of these factors are
degenerated. For example a lower gas density can be compensated by a
higher CR source normalization. According to the CR and diffuse
$\gamma$-ray measurements, the uncertainty of the estimation of
$\gamma$-ray background can be limited to a few tens percent. Here
we adopt a rough estimation of about $20\sim30\%$ uncertainty on
background $\gamma$-rays based on the model fitting of propagation
parameters using B/C ratio data \cite{Lionetto:2005jd}. If we change
the background intensity by $\pm25\%$, the corresponding DM
contribution will change $\sim\pm50\%$ to compensate this variation.
The detectable number of DM subhaloes will also change by $\pm50\%$.

\subsection{Observational uncertainties}

The systematic errors of EGRET are estimated as $\sim 15\%$. If the
statistical errors of the photon counts are taken into account the
errors are larger. It is shown that even the DM contribution varies
$\pm50\%$, the total $\gamma$-ray emission is still consistent with
EGRET observation in $1\sigma$ level. This will correspond to
$\sim50\%$ uncertainty of the detectable number of DM subhaloes.

\section{Conclusion and Discussion}

In this work we study the detection performance of DM subhaloes on
Fermi under the assumption that the ``GeV excess'' discovered by
EGRET comes from the DM annihilation. Considering that EGRET being
the precursor of Fermi it is quite natural to predict the
perspective of Fermi based on the EGRET results. An additional
assumption is that the exceeded part of diffuse $\gamma$-rays at
high Galactic latitude is dominated by the contribution from DM
subhaloes, i.e., DM subhaloes give a large boost factor. Then the
EGRET observation is used as a normalization and we get the
detection perspective of DM subhaloes on Fermi. We find the possible
detectable number of DM subhaloes is from a few tens to several
hundred after 1-year operation of Fermi. For the favored value of
the mass function slope of DM subhaloes $\alpha\approx 1.9$ from
recent highest resolution numerical simulation, there will be
several hundred subhaloes that can be seen.

This prediction has relatively large uncertainties. A major
uncertainty comes from the mass function of DM subhaloes, which is
not well determined by numerical simulations. The mass function
determines the ratio of $\gamma$-ray fluxes between low mass
subhaloes and high mass ones. If the mass function is steep, i.e.,
low mass subhaloes dominate the $\gamma$-ray emission, then the
$\gamma$-ray sky will be more ``diffuse like'' and the detectable
number of DM point sources decreases. Otherwise more DM subhaloes
can be detected. Our result shows that according to the current
knowledge about the mass function from numerical simulations the
detectable number can vary by orders of magnitude. Other
uncertainties from the inner properties of DM subhalo, background
estimation and the observational errors will totally contribute a
factor of $2\sim 3$. Different from previous studies on the DM {\it
indirect search}, our result is independent of the particle physics
property of DM particles.

In spite of the large uncertainties, our result shows that searching
for DM subhaloes may be a promising way for the {\it indirect
search} of DM on Fermi. It is believed that Fermi will open a new
era in the DM study and greatly enrich our knowledge about DM.

Finally we point out that this prediction is based on the
assumption that the EGRET ``GeV excess'' completely comes from the
DM annihilation. If the CR processes contribute a fraction or the
whole of the ``GeV excess'' as proposed in
Ref.\cite{Strong:2004de}, or the ``GeV excess'' is due to the
wrong sensitivity estimation of the detector of EGRET as pointed
out in Ref.\cite{Stecker:2008}, the detectability of DM subhaloes
on Fermi will not be so promising as we have discussed. Especially
the prediction becomes difficult due to the large uncertainties
from the particle physics and distribution of DM.


\begin{thebibliography}{90}

\bibitem{Drees:2000he}
  {Drees M, Kim Y G, Nojiri M M et al}, {Phys. Rev. D}, {2001}, {63}, {035008}

\bibitem{Feng:2005gj}
  {Feng J L, Su S F, Takayama F}, {Phys. Rev.
  Lett.}, {2006}, {96}, {151802}

\bibitem{Baer:2008uu}
  Baer H, Tata X. arXiv:0805.1905 [hep-ph].

\bibitem{Kane:2008gb}
  {Kane G, Watson S}, {Mod. Phys. Lett. A}, {2008}, {23}, {2103}

\bibitem{Bednyakov:1994te}
  {Bednyakov V A, Klapdor-Kleingrothaus H V, Kovalenko S G}, {Phys. Lett. B},
  {1994}, {329}, {5}

\bibitem{Bernabei:1998td}
  {Bernabei R et al. [DAMA Collaboration]}, {Phys. Lett. B},
  {1999}, {450}, {448}

\bibitem{Munoz:2003gx}
  {Munoz C}, {Int. J. Mod. Phys. A}, {2004}, {19}, {3093}

\bibitem{Trotta:2006ew}
  {Trotta R, de Austri R R, Roszkowski L}, {New Astron. Rev.}, 
  {2007}, {51}, {316}

\bibitem{Bouquet:1989sr}
  {Bouquet A, Salati P, Silk J}, {Phys. Rev. D}, {1989}, {40}, {3168}

\bibitem{Bergstrom:2001jj}
  {Bergstrom L, Edsjo J, Ullio P}, {Phys. Rev.
  Lett.}, {2001}, {87}, {251301}

\bibitem{Elsaesser:2004ap}
  {Elsaesser D, MannheimK}, {Phys. Rev. Lett.}, {2005}, {94}, {171302}

\bibitem{deBoer:2006tv}
  {de Boer W, Sander C, Zhukov V et al}, {Phys. Rev. Lett.}, 
  {2005}, {95}, {209001}

\bibitem{Baltz:2001ir}
  {Baltz E A, Edsjo J, Freese K et al}, {Phys. Rev. D}, {2002}, {65}, {063511}

\bibitem{Donato:2003xg}
  {Donato F, Fornengo N, Maurin D et al}, {Phys. Rev. D}, {2004}, {69}, {063501}

\bibitem{Donato:2008yx}
  {Donato F, Fornengo N, Maurin D}, {Phys. Rev. D}, {2008}, {78}, {043506}

\bibitem{Albuquerque:2000rk}
  {Albuquerque I F, Hui L, Kolb E W}, {Phys. Rev.
  D}, {2001}, {64}, {083504}

\bibitem{Barger:2001ur}
  {Barger V D, Halzen F, Hooper D et al}, {Phys. Rev. D}, {2002}, {65}, {075022}

\bibitem{Halzen:2005ar}
  {Halzen F, Hooper D}, {Phys. Rev. D}, {2006}, {73}, {123507}

\bibitem{Barger:2007hj}
  {Barger V D, Keung W Y, Shaughnessy G}, {Phys. Lett. B}, {2008}, {664}, {190}

\bibitem{Gondolo:1999ef}
  {Gondolo P, Silk J}, {Phys. Rev. Lett.}, {1999}, {83}, {1719}

\bibitem{Merritt:2002vj}
  {Merritt D, Milosavljevic M, Verde L et al}, {Phys. Rev. Lett.}, 
  {2002}, {88}, {191301}

\bibitem{Cesarini:2003nr}
  {Cesarini A, Fucito F, Lionetto A et al}, {Astropart. Phys.}, 
  {2004}, {21}, {267}

\bibitem{Dodelson:2007gd}
  {Dodelson S, Hooper D, Serpico P D}, {Phys. Rev.
  D}, {2008}, {77}, {063512}

\bibitem{Bi:2005im}
  {Bi X J}, {Nucl. Phys. B}, {2006}, {741}, {83}

\bibitem{Bi:2007pr}
  {Bi X J}, {Phys. Rev. D}, {2007}, {76}, {123511}

\bibitem{Diemand:2006ik}
  {Diemand J, Kuhlen M, Madau P}, {Astrophys. J.}, {2007}, {657}, {262}

\bibitem{Yuan:2006ju}
  {Yuan Q, Bi X J}, {J. Cosmol. Astropart. Phys.}, {2007}, {0705}, {001}

\bibitem{Pieri:2007ir}
  {Pieri L, Bertone G, Branchini E}, {Mon. Not. Roy. Astron.
  Soc.}, {2008}, {384}, {1627}

\bibitem{Hooper:2003sh}
  {Hooper D, Ferrer F, Boehm C et al}, {Phys. Rev.
  Lett.}, {2004}, {93}, {161302}

\bibitem{Bi:2006av}
  {Bi X J, Hu H B, Zhang X M}, {Eur. Phys. J. C}, {2006}, {48}, {627}

\bibitem{Colafrancesco:2006he}
  {Colafrancesco S, Profumo S, Ullio P}, {Phys. Rev.
  D}, {2007}, {75}, {023513}

\bibitem{Strigari:2006rd}
  {Strigari L E, Koushiappas S M, Bullock J S et al}, {Phys. Rev. D}, 
  {2007}, {75}, {083526}

\bibitem{Bertone:2005xz}
  {Bertone G, Zentner A R, Silk J}, {Phys. Rev. D}, {2005}, {72}, {103517}

\bibitem{Fornasa:2007ap}
  {Fornasa M, Taoso M, Bertone G}, {Phys. Rev. D}, {2007}, {76}, {043517}

\bibitem{Brun:2007tn}
  {Brun P, Bertone G, Lavalle J et al}, {Phys. Rev. D}, {2007}, {76}, {083506}

\bibitem{Wai:2007zza}
  {Wai L L [GLAST-LAT Collaboration]}, {AIP Conf.
  Proc.}, {2007}, {921}, {139}

\bibitem{Hartman:1999fc}
  {Hartman R C et al [EGRET Collaboration]}, {Astrophys. J.
  Suppl.}, {1999}, {123}, {79}

\bibitem{Hunger:1997we}
  {Hunter S D et al [EGRET Collaboration]}, {Astrophys.
  J.}, {1997}, {481}, {205}

\bibitem{Cillis:2005bd}
  {Cillis A N, Hartman R C}, {Astrophys. J.}, {2005}, {621}, {291}

\bibitem{Flix:2004ks}
  {Flix J, Taylor J E, Martinez M et al}, {Astrophys. Space
  Sci.}, {2005}, {297}, {299}

\bibitem{SiegalGaskins:2007dx}
  Siegal-Gaskins J M, Pavlidou V, Olinto A V et al. arXiv:0710.0874

\bibitem{deBoer:2005tm}
  {de Boer W, Sander C, Zhukov V et al}, {Astron.
  Astrophys.}, {2005}, {444}, {51}

\bibitem{deBoer:2004es}
  {de Boer W}, {New Astron. Rev.}, {2005}, {49}, {213}

\bibitem{deBoer:2007sq}
  de Boer W. arXiv:0711.1912

\bibitem{Bi:2006vk}
  {Bi X J, Zhang J, Yuan Q et al}, {Phys. Lett. B}, {2008}, {668}, {87}

\bibitem{Bi:2007ab}
  {Bi X J, Zhang J, Yuan Q}, {Phys. Rev. D}, {2008}, {78}, {043001}

\bibitem{Peirani:2004wy}
  {Peirani S, Mohayaee R, de Freitas P J A}, {Phys. Rev.
  D}, {2004}, {70}, {043503}

\bibitem{Wai:2007px}
  {Wai L [GLAST LAT Collaboration]}, {AIP Conf. Proc.}, {2007}, {903}, {599}

\bibitem{Kuhlen:2008aw}
  {Kuhlen M, Diemand J, Madau P}, {Astrophys. J.}, {2008}, {686}, {262}

\bibitem{Baltz:2008wd}
  {Baltz E A et al}, {J. Cosmol. Astropart. Phys.}, {2008}, {0807}, {013}

\bibitem{Strong:1998pw}
  {Strong A W, Moskalenko I V}, {Astrophys. J.}, {1998}, {509}, {212}

\bibitem{Strong:1998fr}
  {Strong A W, Moskalenko I V, Reimer O}, {Astrophys.
  J.}, {2000}, {537}, {763}

\bibitem{Strong:2004de}
  {Strong A W, Moskalenko I V, Reimer O}, {Astrophys.
  J.}, {2004}, {613}, {962}

\bibitem{Jungman:1995df}
  {Jungman G, Kamionkowski M, Griest K}, {Phys.
  Rept.}, {1996}, {267}, {195}

\bibitem{Gondolo:2004sc}
  {Gondolo P, Edsjo J, Ullio P et al}, {J. Cosmol. Astropart. Phys.}, 
  {2004}, {0407}, {008}

\bibitem{tormen98}
  {Tormen G, Diaferio A, Syer D}, {Mon. Not. Roy. Astron.
  Soc.}, {1998}, {299}, {728}

\bibitem{Moore:1999nt}
  {Moore B, Ghigna S, Governato F et al}, {Astrophys.
  J.}, {1999}, {524}, {L19}

\bibitem{Ghigna:1999sn}
  {Ghigna S, Moore B, Governato F et al}, {Astrophys. J.}, {2000}, {544}, {616}

\bibitem{Zentner:2003yd}
  {Zentner A R, Bullock J S}, {Astrophys. J.}, {2003}, {598}, {49}

\bibitem{DeLucia:2003xe}
  {De Lucia G et al}, {Mon. Not. Roy. Astron. Soc.}, {2004}, {348}, {333}

\bibitem{Diemand:2005vz}
  {Diemand J, Moore B, Stadel J}, {Nature}, {2005}, {433}, {389}

\bibitem{Diemand:2004kx}
  {Diemand J, Moore B, Stadel J}, {Mon. Not. Roy. Astron.
  Soc.}, {2004}, {352}, {535}

\bibitem{Helmi:2002ss}
  {Helmi A, White S D M, Springel V}, {Phys. Rev. D}, {2002}, {66}, {063502}

\bibitem{Gao:2004au}
  {Gao L, White S D M, Jenkins A et al}, {Mon. Not. Roy. Astron.
  Soc.}, {2004}, {355}, {819}

\bibitem{Shaw:2005dy}
  {Shaw L, Weller J, Ostriker J P et al}, {Astrophys.
  J.}, {2006}, {646}, {815}

\bibitem{Madau08}
  {Madau P, Diemand J, Kuhlen M}, {Astrophys. J.}, {2008}, {679}, {1260}

\bibitem{Hofmann:2001bi}
  {Hofmann S, Schwarz D J, Stoecker H}, {Phys. Rev.
  D}, {2001}, {64}, {083507}

\bibitem{Chen:2001jz}
  {Chen X L, Kamionkowski M, Zhang X M}, {Phys. Rev.
  D}, {2001}, {64}, {021302}

\bibitem{Navarro:1996gj}
  {Navarro J F, Frenk C S, White S D M}, {Astrophys. J.}, {1997}, {490}, {493}

\bibitem{Moore:1999gc}
  {Moore B, Quinn T R, Governato F et al}, {Mon. Not. Roy. Astron.
  Soc.}, {1999}, {310}, {1147}

\bibitem{Reed:2003hp}
  {Reed D et al}, {Mon. Not. Roy. Astron. Soc.}, {2005}, {357}, {82}

\bibitem{Diemand:2006ey}
  {Diemand J, Kuhlen M, Madau P}, {Astrophys. J.}, {2006}, {649}, {1}

\bibitem{Bullock:1999he}
  {Bullock J S et al}, {Mon. Not. Roy. Astron. Soc.}, {2001}, {321}, {559}

\bibitem{Yin:2008mv}
  {Yin P F, Liu J, Yuan Q et al}, {Phys. Rev. D}, {2008}, {78}, {065027}

\bibitem{Berezinsky:1992mx}
  {Berezinsky V S, Gurevich A V, Zybin K P}, {Phys. Lett.
  B}, {1992}, {294}, {221}

\bibitem{Lavalle:1900wn}
  {Lavalle J, Yuan Q, Maurin D et al}, {Astron.
  Astrophys.}, {2008}, {479}, {427}

\bibitem{Morselli:2002cc}
  {Morselli A}, {Int. J. Mod. Phys. A}, {2002}, {17}, {1829}

\bibitem{Eke:2000av}
  {Eke V R, Navarro J F, Steinmetz M}, {Astrophys.
  J.}, {2001}, {554}, {114}

\bibitem{Lionetto:2005jd}
  {Lionetto A M, Morselli A, Zdravkovic V}, {J. Cosmol. Astropart.
  Phys.}, {2005}, {0509}, {010}

\bibitem{Stecker:2008}
  {Stecker F W, Hunter S D, Kniffen D A}, {Astropart.
  Phys.}, {2008}, {29}, {25}

\end{thebibliography}
\end{document}